\documentstyle[preprint, aps]{revtex}

\draft
\tighten

\begin{document}
\title{Quark mass density- and temperature- dependent model for strange quark matter}
\author{Yun Zhang$^1$, Ru-Keng Su$^{2,1}$, Shuqian Ying$^{1}$, and Ping Wang$^{1,3 }$}
\address{$^1$Department of physics, Fudan University, Shanghai 200433, P. R. China\\
$^2$China Center of Advanced Science and Technology (World Laboratory),\\
P. O. Box 8730, Beijing 100080, P. R. China \\
$^3$Institute of High Energy Physics, P. O. Box 918, Beijing 100080, P. R.
China}
\maketitle

\begin{abstract}
It is found that the radius of a stable strangelet decreases as the
temperature increases in a quark mass density-dependent model. To overcome
this difficulty, we extend this model to a quark mass density- and
temperature- dependent model in which the vacuum energy density at zero
baryon density limit $B$ depends on temperature. An {\em ansatz} which reads 
$B=B_0[1-a({T/T_c})+b({T/T_c})^2]$ is introduced and the regions for the
best choice of the parameters are studied.
\end{abstract}


It has been expected that a new state of matter, namely, the quark-gluon
plasma (QGP) can be formed in the relativistic heavy-ion collision. Many
signatures of the formation of QGP, such as $J/\Psi $ suppression,
strangeness enhancement, thermal dilepton and electromagnetic radiation etc.
have been suggested \cite{ins1}. A lot of them have been found in recent
experiments as pointed by CERN collaboration. However whether QGP has been
got is still an open question, because many signatures suggested can also be
explained by other treatments \cite{ins2}. To search an unambiguous
signature is an essential task for the study of QGP. Due to this reason, the
strangelets, the strange quark matter (SQM) in finite lumps, have attracted
much attention in recent years. Greiner and his coworkers \cite{ins3} argued
that strangelets might be produced in ultrarelativistic heavy-ion collisions
and could serve as an unambiguous signature for the formation of QGP.

Since the speculation of Witten \cite{1} that the strange quark matter (SQM)
would be more stable than the normal nuclear matter and nuclei, many models
including MIT bag model \cite{2,3,4}, quark-meson coupling model \cite{5},
quark mass-density dependent model (QMDD) \cite{7,8,9,10,peng1,peng2,wang},
Friedberg-Lee model \cite{11}, chiral SU(3) quark model \cite{12} and etc. 
\cite{13} have been employed to predict the behavior of SQM in the bulk. It
is of interest to investigate whether above models which are successful for
bulky strange matter can be used to describe strangelets.

This paper evolves from an attempt to study this problem for the QMDD model.
This model can provide a dynamical description of the confinement mechanism
and explain the stability of SQM successfully via the suggestion of a
density dependent masses for u, d and s quarks. It is found that a
difficulty emerges if the QMDD\ model is used to describe strangelets at
finite temperature. The radius of strangelet decreases as the temperature
increases. This is of course unreasonable. Since the mass of hadrons is
observed to be dependent on temperature, we extend the QMDD model to a quark
mass density- and temperature-dependent (QMDTD) model, and show that after a
suitable choice of the adjusted parameters in the function of $B(T)$, the
radius of the strangelet increases with the rise of temperature.

According to the QMDD model, the masses of u, d quarks and strange quarks
(and the corresponding anti-quarks) are given by \cite{7,8}

\begin{eqnarray}
m_q &=&{\frac B{3\tilde{n}_B}},\hspace{0.8cm}(q=u,d,\bar{u},\bar{d}),
\label{1} \\
m_{s,\bar{s}} &=&m_{s0}+{\frac B{3\tilde{n}_B}},  \label{2}
\end{eqnarray}
where $\tilde{n}_B$ is the baryon number density, $m_{s0}$ is the current
mass of the strange quark and $B$ is the vacuum energy density inside the
bag.

The thermodynamic potential is

\begin{equation}
\Omega =\sum_i\Omega _i=-\sum_iT\int_0^\infty dk{\frac{dN_i}{dk}}\ln \left(
1+e^{-\beta (\varepsilon _i(k)-\mu _i)}\right) ,  \label{3}
\end{equation}
where $i$ stands for $u,d,s$ (or $\bar{u},\bar{d},\bar{s}$ ) and the
electron $e$($e^{+}$), $\mu _i$ is the corresponding chemical potential.
Since the masses of quarks depend on density, the thermodynamic potential $%
\Omega $ is not only a function of temperature, volume and chemical
potential, but also of density. How to treat the thermodynamics with
density-dependent particle masses self-consistently is a serious problem and
has made many wrangles in the references \cite{8,9,peng1,wang}. Here after
we follow the treatment of ref.\cite{peng1}. The number density $\tilde{n}_i$%
, the total pressure $p$ and the total energy density $\varepsilon $ are
given by

\begin{equation}
\tilde{n}_i=-{\frac 1V}\left. {\frac{\partial \Omega }{\partial \mu _i}}%
\right| _{T,\tilde{n}_B},  \label{4}
\end{equation}

\begin{equation}
p=-{\frac 1V}\left. {\frac{\partial (\Omega /\tilde{n}_B)}{\partial (1/%
\tilde{n}_B)}}\right| _{T,\mu _i}=-{\frac \Omega V}+{\frac{\tilde{n}_B}V}%
\left. {\frac{\partial \Omega }{\partial \tilde{n}_B}}\right| _{T,\mu _i},
\label{5}
\end{equation}

\begin{equation}
\varepsilon ={\frac \Omega V}+\sum_i\mu _i\tilde{n}_i-{\frac TV}\left. {%
\frac{\partial \Omega }{\partial T}}\right| _{\mu _i,\tilde{n}_B},  \label{6}
\end{equation}
respectively. The baryon number density $\tilde{n}_B$ reads

\begin{equation}
\tilde{n}_B={\frac 13}(\tilde{n}_u+\tilde{n}_d+\tilde{n}_s)\hspace{0in}.
\label{7}
\end{equation}
The conditions of charge neutrality and chemical equilibrium are 
\begin{eqnarray}
2\tilde{n}_u &=&\tilde{n}_d\hspace{0in}+\tilde{n}_s+3\tilde{n}_e,  \label{8}
\\
\mu _s &=&\mu _d,\hspace{0.5cm}\mu _s=\mu _u+\mu _e.  \label{9}
\end{eqnarray}

Now we are in a position to use the above formulation to investigate the
strangelets. Instead of one of plane wave, the density of states $%
{\displaystyle {dN_i \over dk}}%
$ for a sphere with radius $R$ is needed in our calculations. It was given
by a multi-reflection theory \cite{15}

\begin{equation}
\frac{dN_i}{dk}=g_i\left[ \frac{k^2V}{2\pi ^2}+f_s^{(i)}\left( \frac{m_i}k%
\right) kS+f_c^{(i)}\left( \frac{m_i}k\right) C+...\right] ,  \label{10}
\end{equation}
where $V=\frac 43\pi R^3,S=4\pi R^2,C=8\pi R,g_i=6$ for quarks and
antiquarks, $g_i=2$ for $e$ and $e^{+}$.The second term on the right hand
side of Eq.(\ref{10}) corresponds to the surface contribution. It is shown 
\cite{17}

\begin{equation}
f_s^{(i)}\left( \frac{m_i}k\right) =-\frac 1{8\pi }\left( 1-\frac 2\pi
\arctan \frac k{m_i}\right) .  \label{11}
\end{equation}
This term is zero for massless quarks. The third term on the right hand side
of Eq.(\ref{10}) comes from curvature of the bag surface. It can not be
obtained by the multi-reflection theory directly except for two limiting
cases $m_i\rightarrow 0$ and $m_i\rightarrow \infty $. Madsen proposed\cite
{18}

\begin{equation}
f_c^{(i)}\left( \frac{m_i}k\right) =\frac 1{12\pi ^2}\left( 1-\frac{3k}{2m_i}%
\left( \frac \pi 2-\arctan \frac k{m_i}\right) \right) .  \label{12}
\end{equation}
This simple form agrees to the above mentioned two values in the
corresponding limits.

For the strangelets, the stability condition reads

\begin{equation}
\frac{\partial F}{\partial R}=0,  \label{13}
\end{equation}
where $F$ is the total free energy. And it can be given by 
\begin{equation}
F=E-TS,  \label{13-1}
\end{equation}
where $E=\varepsilon V$ is the total energy, 
\begin{equation}
S=\sum_iS_i=-\sum_i\int_0^\infty dk{\frac{dN_i}{dk}}\left[ n_i\ln
n_i+(1-n_i)\ln (1-n_i)\right]   \label{13-2}
\end{equation}
is the entropy, $n_i$ is the distributing function of fermions and $i$
stands for $u,d,s$ (or $\bar{u},\bar{d},\bar{s}$) quarks and the electron $e$%
($e^{+}$). Substituting Eqs.(\ref{10}), (\ref{11}), (\ref{12}) into Eq.(\ref
{3}) and using Eq.(\ref{13}), we can obtain the stable radius $R$ of the
strangelet, which is a function of temperature.

The curves for $F$ per baryon number $A$ vs. $R$ of QMDD\ model at zero
temperature and at $T=$ $50\mbox{MeV}$ are shown in Fig.1 by solid line and
dashed line, respectively. The values of $A$, $B$, $m_{s0\mbox{
}}$are chosen as 
\begin{equation}
A=20,B=170\mbox{MeV}\mbox{fm}^{-3},m_{s0}=150\mbox{MeV}.  \label{13-a}
\end{equation}
It is clear that the radius for minimum $F$ decreases as temperature
increases. The temperature dependence on radius $R$ is displayed in Fig.2,
in which it can be seen that $R(T)$ is a monotonously decreasing function.
This is of course unreasonable because the bag is expected to expand as the
increase of temperature.

To understand this result physically, let us recall what happens for the
masses of nucleons and mesons when temperature increases. In fact, the
effective masses of nucleons, effective masses and screening masses of
mesons, are all dependent on temperature \cite{a1,a2,a3}.We can sum the
tadpole diagrams and the exchange diagrams for nucleons, the vacuum
polarization diagrams for mesons by Thermo Field Dynamics and find that the
masses of nucleons and mesons all decrease as temperature increases. This
result for $\rho -$meson is in agreement with recent experiments\cite{a5}.
According to the constituent quark model, the nucleon is constructed by
three quarks and the meson by two quarks. This means that to satisfy quark
model we must consider the temperature dependence of the quark mass. But
this effect has not been taken into account in Eqs.(\ref{1}) and (\ref{2})
if $B$ is a constant. Therefore we come to a conclusion that the unphysical
result of QMDD model in studying strangelets comes from the assumption that $%
B$ is temperature independent.

Although it is possible to find the function $B(T)$ from a calculations of
the vacuum energy, but in this paper, instead of the vacuum energy
calculation, we study this problem more generally by introducing an {\em \
ansatz}, namely

\begin{equation}
B=B_0\left[ 1-a\left( \frac T{T_c}\right) +b\left( \frac T{T_c}\right)
^2\right] ,  \label{15}
\end{equation}
where $a$, $b$ are two adjust parameters, and $T_{c\mbox{
}}=170\mbox{MeV}$ is the critical temperature of the quark deconfinement
phase transition. The reasons for our choice are: first, the calculation of $%
B(T)$ are model-dependent and the results given by different model are very
different \cite{20}\cite{21}; secondly, almost all previous calculations are
based on MIT bag model, but now we discuss the QMDD model. In fact we can
imagine that the Eq.(\ref{15}) is the temperature expansion of $B$ in the
low temperature regions. Since $B$ is zero when $T=T_c$, a condition 
\begin{equation}
1-a+b=0  \label{16}
\end{equation}
is imposed and only one parameter $a$ can be adjusted.

Introducing Eqs.(\ref{15}) and (\ref{16}), we extend the QMDD\ model to a
quark mass density- and temperature- model(QMDTD). The results of our model
obtained by substituting Eqs.(\ref{15}) and (\ref{16}) into Eqs.(\ref{1})
and (\ref{2}) are shown in Figs.(1), (3) and (4).The $F/A$ vs. stable $R$
curve for $T=50\mbox{MeV}$, $a=0.65$ given by QMDTD model is shown in
Fig.(1) by dot line. We see that the value of stable radius increases from $%
R_{(T=0)}=2.27\mbox{fm}$ to $R_{(T=50\mbox{MeV})}=2.31\mbox{fm}$, but
decreases to $R_{(T=50\mbox{MeV})}=2.18\mbox{fm}$ for QMDD model as shown by
dashed-line.

The value of parameter $a$ can affect the result significantly. The range of
possible $a$ is determined by physical constraints. For example, there are
at least two physical constraints: (1), the stable radius $R$ should
increase with temperature; (2), the energy per baryon $E/A$ increases with
temperature also. To show the importance of first constraint, we plot the $%
R(T)$ curves for $a=$ $-0.20$, $0.20$, $0.40$, $0.65$, $0.80$ in Fig.(3),
respectively. We see that $R(T)$ becomes a monotonously increasing function
in the regions $0\leq T\leq 80\mbox{MeV}$ when $a$ $\geq 0.65$. On the other
hand, the $E/A$ vs. $T$ curves for $a=1.50$, $0.90$, $0.80$, $0.65$, $0.20$
are shown in Fig.(4) to investigate the relevance of the second constraint.
We find $E/A$ vs. $T$ curves becomes monotonously increasing function in the
same temperature region when $a\leq 0.8$. Therefore the best values for
parameter $a$ are in the range

\begin{equation}
0.65\leq a\leq 0.8.  \label{17}
\end{equation}

In summary, in order to overcome the difficulty related to the reduction of $%
R$ with $T$, we suggested a QMDTD model. An ansatz for the temperature
dependence of vacuum energy $B$ (Eq.(\ref{15})) was introduced. We found
that the parameters $a$ must lies in the range $0.65\leq a\leq 0.8$. Our
model can be used to study strangelet, in addition it can also be employed
to address systematically the properties of SQM in bulk.

This work was supported in part by the NNSF of China under construct
No.19975010, No.19875009, and the Foundation of Education Ministry of China.

\section{Figure Captions}

Fig.1 The total free energy per baryon $F/A$ as a function of radius $R$,
for temperature $T=0$(solid line), $T=50\mbox{MeV}$ in QMDD model (dashed
line) and $T=50\mbox{MeV}$, $a=0.65$ in QMDTD\ model (dotted line).

Fig.2 The radius $R$ as a function of temperature $T$ for QMDD\ model.

Fig.3 The radius $R$ as a function of temperature $T$ for QMDTD\ model with
various values for the parameter $a$ as indicated.

Fig.4 The energy per baryon $E/A$ as a function of temperature $T$ for QMDTD
model with various values for the parameter $a$ as indicated.

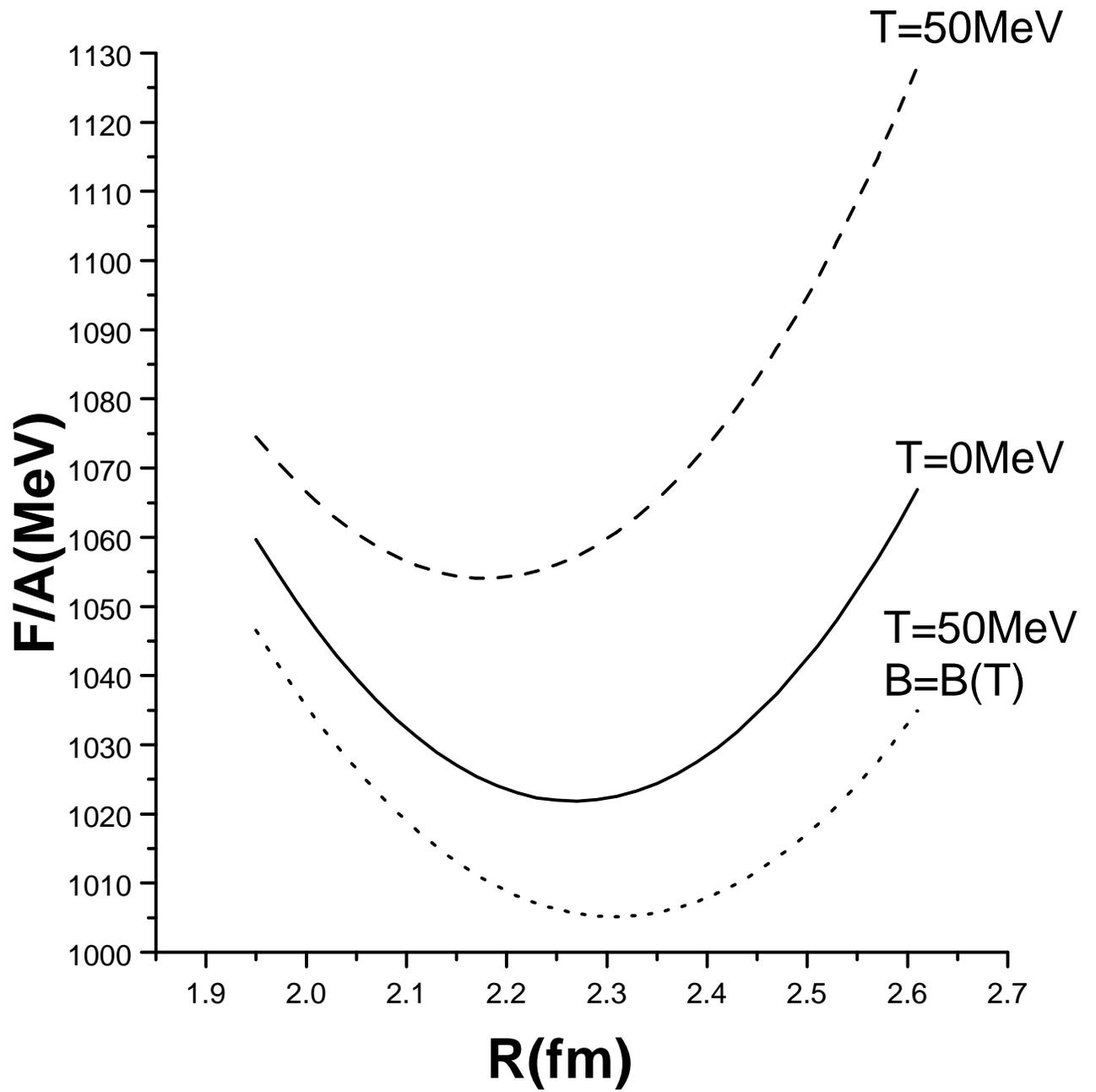
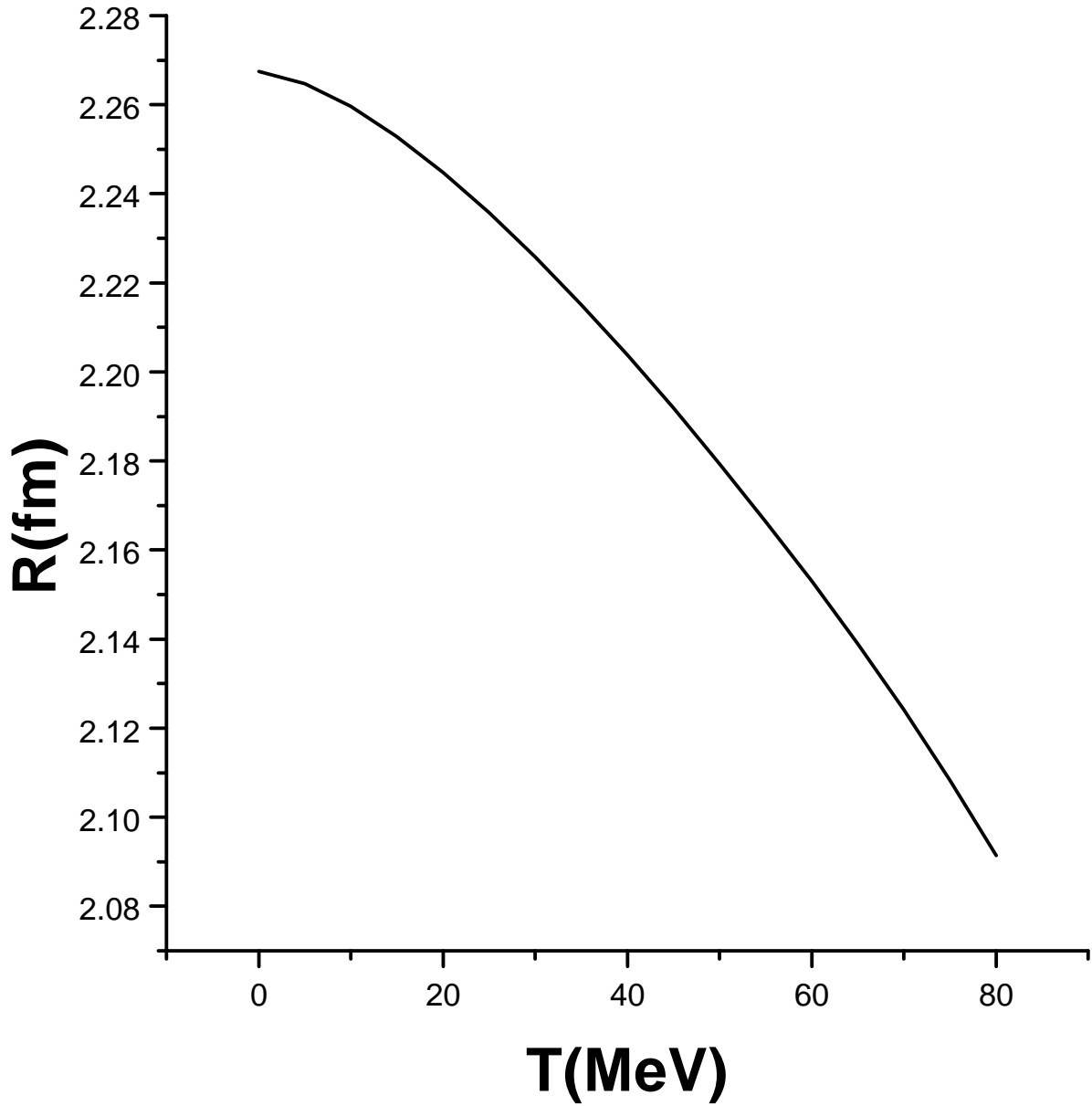
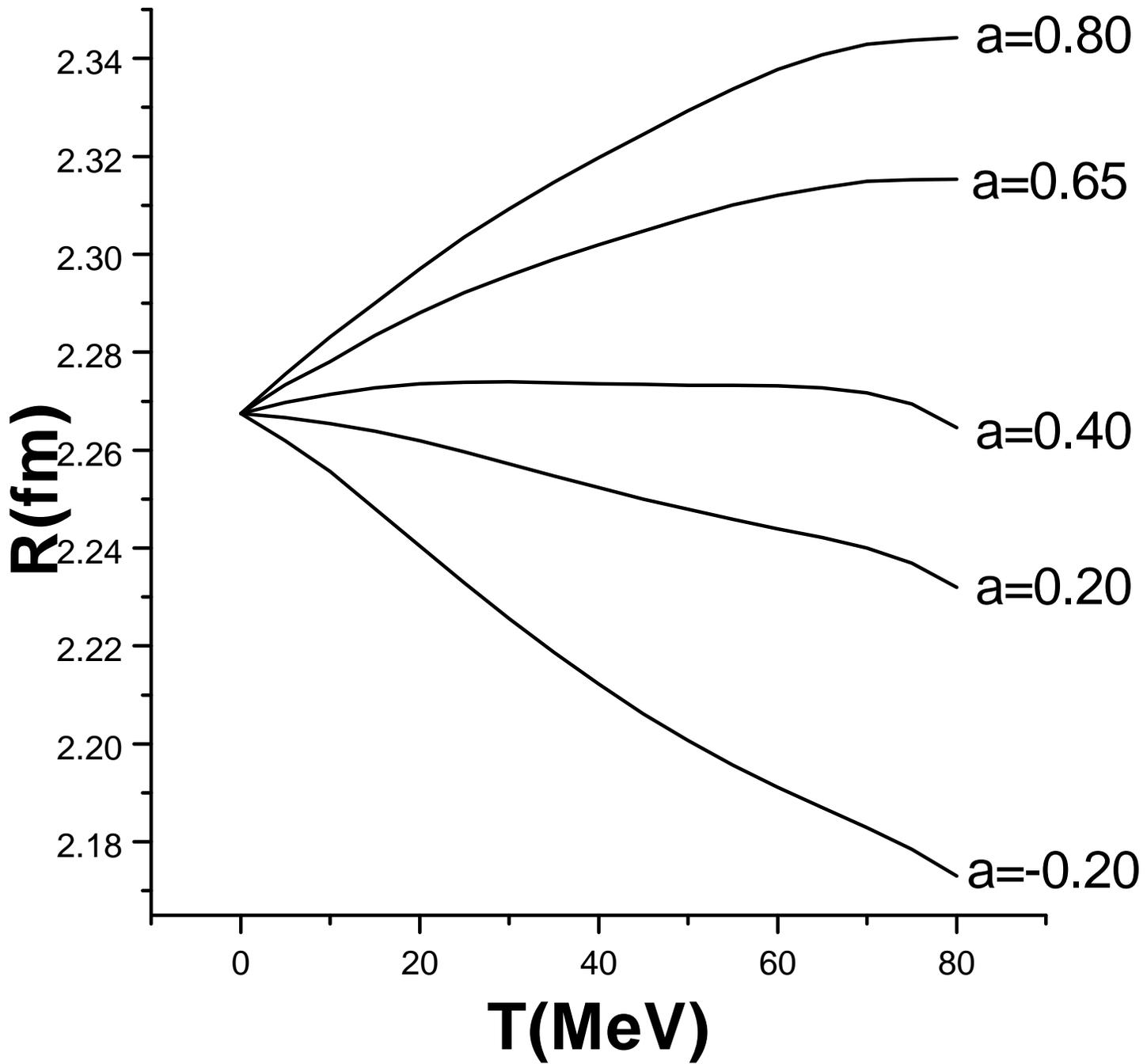
\end{document}